\documentclass[conference]{sig-alternate-05-2015}

\usepackage{alltt                                    % I like these
          , multirow
          , booktabs
          , listings
          , graphicx
          ,float
	,cite
          ,verbatim
         ,mathtools
	,url
	,amsmath
}
\usepackage[table]{xcolor}
\usepackage[numbers]{natbib}     % this is a better citation system
\usepackage{syntax}
\usepackage{algorithm}
\usepackage{enumitem}
\usepackage{threeparttable}
\usepackage{framed}
\usepackage{algpseudocode}
\usepackage{hhline}
\usepackage{balance}
\usepackage{setspace}
\usepackage{afterpage}

\usepackage{expl3}
\ExplSyntaxOn
\newcommand\latinabbrev[1]{
  \peek_meaning:NTF . {% Same as \@ifnextchar
    #1\@}%
  { \peek_catcode:NTF a {% Check whether next char has same catcode as \'a, i.e., is a letter
      #1., \@ }%
    {#1., \@}}}
\ExplSyntaxOff

\newcommand{\CASE}[1]{\STATE \textbf{case} #1\textbf{:} \begin{ALC@g}}
\newcommand{\ENDCASE}{\end{ALC@g}}

\newcommand{\DEFAULT}{\STATE \textbf{default:} \begin{ALC@g}}
\newcommand{\ENDDEFAULT}{\end{ALC@g}}
\newcommand{\DEFAULTLINE}[1]{\STATE \textbf{default:} }
\let\footnotesize\scriptsize

\algnewcommand{\LineComment}[1]{\State \(\triangleright\) #1}

\newlength{\oldtextfloatsep}

\renewcommand{\refname}{9.~~~ REFERENCES}

\newsavebox{\supbox}% Superscript box
\newcommand{\bsup}{\begin{lrbox}{\supbox}$\tt\scriptstyle}% Superscript begin
\newcommand{\esup}{$\end{lrbox}{}^{\usebox{\supbox}}}% Superscript end
\def\eg{\latinabbrev{e.g}}
\def\ie{\latinabbrev{i.e}}

\definecolor{lightpurple}{rgb}{0.8,0.8,1}
\definecolor{codebg}{RGB}{255,255,255}
\definecolor{commentcolor}{RGB}{11,140,11}
\begin{document}
\toappear{}

\setcopyright{acmcopyright}
\isbn{978-1-4503-4205-6}

\conferenceinfo{ASE}{'2016 Singapore}

%\title{CORRECT: Code Reviewer Recommendation for Github Pull Requests using Cross-Project and Technology Experience}
%\title{CORRECT: Reviewer Recommendation for GitHub Pull Requests using Cross-Project and Technology Experience}
\title{CORRECT: Code Reviewer Recommendation at GitHub for Vendasta Technologies}

%\author{\IEEEauthorblockN{Mohammad Masudur Rahman  ~~~ Chanchal K. Roy ~~~ $^\dagger$Jason Collins}
%\IEEEauthorblockA{University of Saskatchewan, Canada,  $^\dagger$Google Inc., USA\\
%\{masud.rahman, chanchal.roy\}@usask.ca,  $^\dagger$jasonco@google.com}
%}

\numberofauthors{1} %  in this sample file, there are a *total*
% of EIGHT authors. SIX appear on the 'first-page' (for formatting
% reasons) and the remaining two appear in the \additionalauthors section.
%
\author{
% You can go ahead and credit any number of authors here,
% e.g. one 'row of three' or two rows (consisting of one row of three
% and a second row of one, two or three).
%
% The command \alignauthor (no curly braces needed) should
% precede each author name, affiliation/snail-mail address and
% e-mail address. Additionally, tag each line of
% affiliation/address with \affaddr, and tag the
% e-mail address with \email.
%
% 1st. author
\alignauthor
Mohammad Masudur Rahman$^\star$~~~Chanchal K. Roy$^\star$~~~Jesse Redl$^\S$~~~Jason A. Collins$^\dagger$\\
       \affaddr{University of Saskatchewan$^\star$, Canada, Vendasta Technologies$^\S$, Canada, Google Inc.$^\dagger$, USA}\\
      \email{\{masud.rahman, chanchal.roy\}@usask.ca$^\star$, jredl@vendasta.com$^\S$, jasonco@google.com$^\dagger$}
% 2nd. author
%\alignauthor
%Chanchal K. Roy\\
%       \affaddr{University of Saskatchewan, Canada}\\
%\affaddr{chanchal.roy@usask.ca}
%% 3rd. author
%\alignauthor
%$^\dagger$Jason Collins\\
%       \affaddr{$^\dagger$Google Inc., USA}\\
%\affaddr{$^\dagger$jasonco@google.com}
% 3rd. author
}

\CopyrightYear{2016} 
\setcopyright{acmcopyright}
\conferenceinfo{ASE'16,}{September 03-07, 2016, Singapore, Singapore}
\isbn{978-1-4503-3845-5/16/09}\acmPrice{\$15.00}
\doi{http://dx.doi.org/10.1145/2970276.2970283}

\maketitle

\begin{abstract}
%Code review is considered as one the most important developer collaborations in modern collaborative software development. 
Peer code review locates common coding standard violations and simple logical errors in the early phases of software development, and thus, reduces overall cost. 
Unfortunately, at GitHub, identifying an appropriate code reviewer for a pull request is challenging given that reliable information for reviewer identification is often not readily available. 
In this paper, we propose a code reviewer recommendation tool--CORRECT--that considers not only the relevant cross-project work experience (\eg\ external library experience) of a developer but also 
her experience in certain specialized technologies (\eg\ Google App Engine) associated with a pull request
for determining her expertise as a potential code reviewer. We design our tool using client-server architecture, and then package the solution as a Google Chrome plug-in. 
Once the developer initiates a new pull request at GitHub, our tool automatically analyzes the request, mines two relevant histories, and then returns a ranked list of appropriate code reviewers for the request within the browser's context.
\noindent
Demo: https://www.youtube.com/watch?v=rXU1wTD6QQ0

\end{abstract}

% A category with the (minimum) three required fields
%\category{H.4}{Information Systems Applications}{Miscellaneous}
%A category including the fourth, optional field follows...
%\category{D.2.8}{Software Engineering}{Heuristics}[cross-project experience, specialized technology experience]

%<ccs2012>
%<concept>
%<concept_id>10011007.10011006</concept_id>
%<concept_desc>Software and its engineering~Software notations and tools</concept_desc>
%<concept_significance>500</concept_significance>
%</concept>
%<concept>
%<concept_id>10011007.10011074.10011134.10011135</concept_id>
%<concept_desc>Software and its engineering~Programming teams</concept_desc>
%<concept_significance>500</concept_significance>
%</concept>
%</ccs2012>



%\terms{Theory}

%\begin{keywords}
%Code reviewer recommendation, cross-project experience, technology experience, GitHub, pull request  
%\end{keywords}

\section{Introduction}
%Software development practices have dramatically changed over the last decade, and software projects are now developed not only in a collaborative environment but also in a distributed fashion \cite{seafood}.
Peer code review is reported to be highly effective for locating coding standard violations or for performing simple logical verifications \cite{vcc,expect,contempo}.
It also helps identify source code issues (\eg\ vulnerabilities) in the early phases of development, and thus, reduces overall cost for the software project \cite{vcc,expect}.
GitHub promotes a distributed and collaborative software development through pull requests and code reviews respectively.
In GitHub, a developer forks from an existing repository (\ie\ project), works on certain module of her interest, and then submits the changed files to the repository using a pull request \cite{pull}. 
During pull request submission, the developer is expected to choose one or more code reviewers who would review the code carefully before accepting the changes as a contribution.
%The administrators of that repository then review the changes, consults with the developer, and accepts the contribution if satisfied.
Unfortunately, choosing an appropriate reviewer for a pull request is a significant challenge \cite{reduce}, and to date, GitHub does not provide any support for this.
Reliable information on reviewers' expertise (\eg\ technology skill) is often not readily available, and it needs to be carefully mined from the codebase.
%Thus, reviewer identification task is even more challenging and time-consuming for the novice developers who are less familiar with the codebase as well as the skills of the hundreds of fellow developers.
Thus, the task of identifying appropriate reviewers is even more challenging and time-consuming for novice developers since they are neither familiar well with the codebase nor are aware of the skills of the hundreds of fellow potential reviewers.
%They often experience difficulties in identifying the code reviewers from hundreds of project members.
Such challenge is prevalent not only in open source development but also in
the industrial environment where a company (\eg\ Vendasta Technologies) strives to maintain code quality in the commercial software development and encourages collaborations among the developers in the form of peer code reviews.
%New developers working on a project 
%Fortunately, an automated support in identifying the appropriate developers for code review can greatly help in this regard.

Fortunately, there have been several studies that recommend code reviewers by analyzing past code review history (\eg\ line change history \cite{reduce}, review comments \cite{yu,xin}), project directory structure \cite{pick, rveffect}, and developer collaboration network \cite{yu}.  
%Similarly, studies on expert recommendation for software bugs also exploit different software artifacts \cite{kevic,authorship} and developer communication history \cite{sna}.
%version control history \cite{kevic, bosu}, authors' meta data \cite{authorship, discovery}, and developer communication history \cite{brazil, seafood, sna, ghadeer}, and perform static analysis such as code similarity analysis \cite{ghadeer}.
%In short, 
In short, the existing studies mostly rely on the work history of a developer (\ie\ potential reviewer) within a particular project and her collaboration history with other developers for determining her expertise.  
However, no studies consider the cross-project experience or the experience in various specialized technologies of a developer, and thus, they fall short in handling certain challenges.
First, in industry, software developers often reuse software components (\eg\ libraries) that were previously developed by themselves for cost-effective and faster development.
Thus, their contributions scatter throughout different projects in the code repositories of the organization. Such contributions are a great proxy to their experience. 
Unfortunately, the existing studies on code reviewer recommendation completely ignore such information in expertise determination, and their recommendations are merely based on the contribution details within a particular project.
Second, underlying tools and technologies of software projects are rapidly changing, and modern projects often involve different  specialized and cutting edge technologies such as \texttt{map-reduce, task queues, urlfetch, memcache} and \texttt{pipeline}.
Hence, code reviewers for a pull request are expected to have expertise in such technologies.
However, neither mining of the revision history of changed files nor mining of the developer collaboration history, as the existing studies do, might be sufficient enough to ensure that.
%One possible way is to analyze the past experience of a developer with those technologies.
%an analysis on the cross-project work experience of the developers is essential to determine thier relevant expertise for a pull request.
%Third, code reviewer recommendation entirely based on version history analysis might not be effective in an organizational context if the recommended developers are unavailable (\ie\ already left the organization) or overloaded with review requests.
%Third, recently proposed tools are reported as effective in identifying simple coding rule violations \cite{reduce, panichella}, and thus, human reviewers are expected to identify potential logical concerns such as security vulnerabilities \cite{vcc} or performance bottlenecks \cite{expect}.
%Such identifications often require special skills or application domain expertise, and previous developers of the changed files might always not be the most appropriate candidates for a code review. 
Thus, a technique that can analyze both relevant cross-project experience and specialized technology experience of a developer for a pull request, is likely to overcome the above challenges.

In this paper, we present a novel code reviewer recommendation tool--CORRECT-- 
%(\textbf{Co}de \textbf{R}eviewer \textbf{Re}commen-dation based on \textbf{C}ross-project and \textbf{T}echnology experience),
for pull requests at GitHub. 
The tool determines eligibility of a developer as code reviewer for a pull request by analyzing her past work experience with (1) external software libraries and (2) specialized technologies used by the pull request.
Reference to the external libraries (\ie\ software units external to the working project) in the code generally suggests one's working experience with such libraries, and we call it \emph{cross-project experience}.
%\FrameSep3pt
%\setlength{\topsep}{2pt}%
%\begin{framed}
Our key idea is-- \emph{if a past pull request uses similar external software libraries or similar specialized technologies to the current pull request, then the past request is relevant to the current request, and thus, its reviewers are also potential candidates for the code review of the current request}.
%\end{framed}

We first mine the external library and technology information from current pull request using static analysis, and then identify the relevant (\ie\ similar) requests in terms of library and technology similarities from the recently submitted pull request collection.
%Since such similarity measures estimate the relevance between pull requests, we propagate those measures to corresponding reviewers for heuristically capturing their expertise for the current pull re
We then propagate the similarity score for each relevant request to its corresponding code reviewers as a proxy to the shared experience in external libraries and specialized technologies with the current request.
Thus, each of the candidates accumulates scores for all relevant requests, and finally, the technique returns a ranked list of code reviewers.
%To the best of our knowledge, ours is the first study that exploits the benefits of using cross-project experience and specialized technology experience (of the developers) in recommending code reviewers for a pull request.
%technology experience or cross-project experience as an expertise dimension despite of their potential for code reviewer recommendation.
We adopt a client-server architecture for our recommendation system where the client module is packaged as a Google Chrome plug-in (\ie\ as per the specification of Vendasta Technologies), and the server module is hosted as a web service. 
%Both modules are available online \cite{correct} for replication or third party use.
%Thus, we make the following contributions in the paper:
To summarize, our proposed tool provides the following features to support Vendasta software developers in the selection of appropriate code reviewers:
\begin{enumerate}[noitemsep,topsep=1pt]
\item automatically analyzes the technical details of a given pull request, and recommends a ranked list of appropriate reviewers for its code review,
\item automatically captures and leverages two expertise dimensions of a developer--\emph{cross-project experience} and \emph{specialized technology experience}--for determining her expertise/eligibility as a code reviewer,
\item offers customized recommendations for the developers using open authentication of GitHub,
\item complements the existing pull request submission utility of GitHub through a Google Chrome plug-in, and 
\item provides a client-server architecture for seamless integration of our code reviewer recommendation service.
\end{enumerate}

While this paper focuses on the tool aspect of our code reviewer recommendation approach, we refer the readers to the original paper \cite{correct} for further details.

\begin{figure}[!t]
\centering
\includegraphics[width=3.3in]{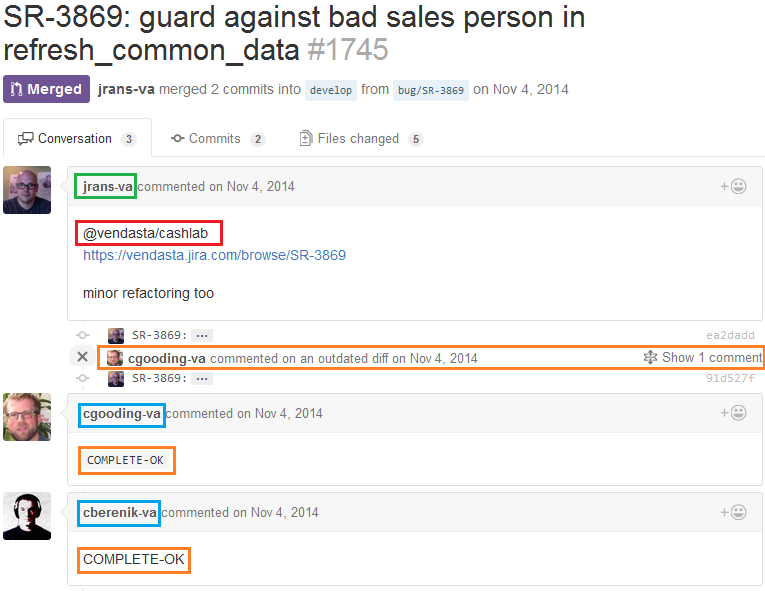}
\vspace{-.6cm}
\caption{Code review interface at GitHub}
\vspace{-.5cm}
\label{fig:mcr}
\end{figure}

\begin{figure*}[!t]
\centering
\includegraphics[width=5.5in]{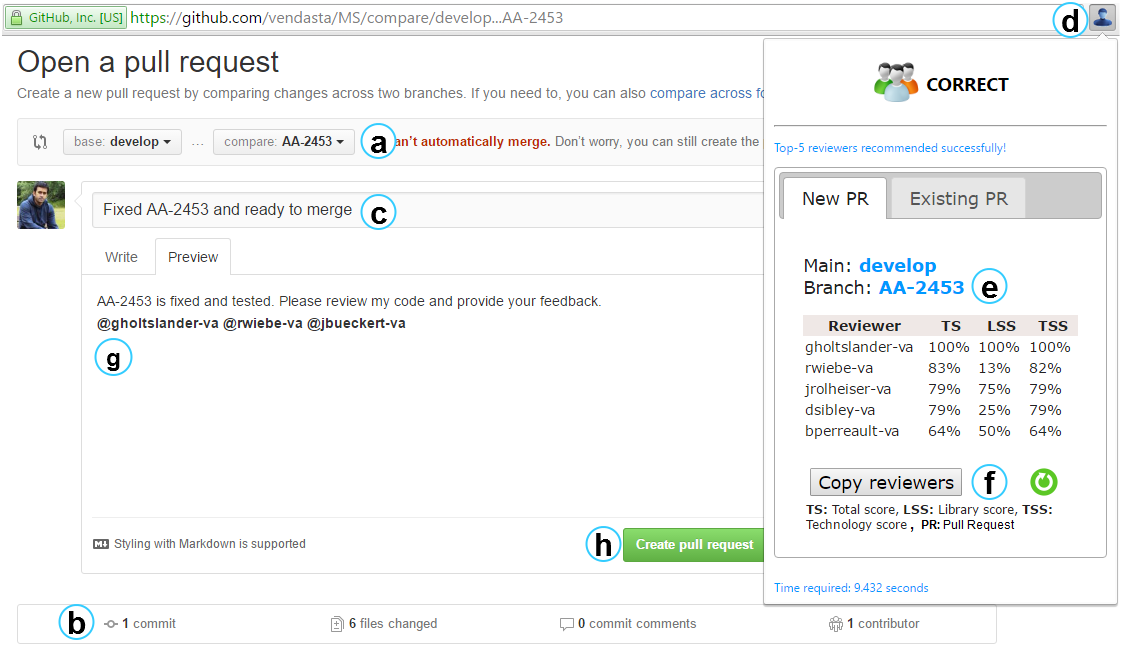}
\vspace{-.3cm}
\caption{User interface of CORRECT Tool}
\vspace{-.4cm}
\label{fig:ui}
\end{figure*}

\vspace{-.1cm}
\section{Modern Code Review (MCR)}\label{sec:mcr}
\emph{Code review} refers to a manual assessment of source code that identifies potential defects (\eg\ logical errors) and quality problems (\eg\ coding standard violations) in the code \cite{modern}. 
%There exist different ways for code review including the most formal \emph{code inspection} to the least formal \emph{over-the-shoulder} review \cite{reduce}.
In recent years, code review has been assisted with various tools, which is less formal and more popular than the traditional review techniques \cite{reduce}. Such code review is called \emph{Modern Code Review} (MCR) \cite{modern}. 
It is widely adopted both by the commercial organizations (\eg\ Microsoft) and by the open-source communities (\eg\  Android, LibreOffice).
%Existing code review tools such as \emph{ReviewBot} \cite{reduce} and \emph{RevFinder} \cite{pick} are mostly based on Gerrit, a web-based code review system.
%\footnote{Web based code review and project management for Git based projects}.
%GitHub also provides a similar feature for conducting code review by human developers through pull requests where a developer can request her peers for code review during a pull request submission.  
As an example, developer \emph{jrans-va} (\ie\ green box, Fig. \ref{fig:mcr}) requests a development team-- \emph{cashlab} (\ie\ red box, Fig. \ref{fig:mcr})-- for code review during the submission of the pull request \#1745. The request contains two commits associated with five changed source files.
Two developers-- \emph{cgooding-va} and \emph{cberenik-va} (\ie\ blue boxes)-- from the team analyze the commits from the pull request, perform the code review, and then post their feedback using comments (\ie\ orange boxes, Fig. \ref{fig:mcr}). 
%The submitter then addresses the comments, and resubmits the changed code. 
%Once the reviewers are satisfied and add the \emph{COMPLETE-OK} label to the reviewed code (\ie\ orange boxes, Fig. \ref{fig:mcr}), the pull request is merged into the codebase.
Unfortunately, despite assistance from the static analysis tools \cite{reduce}, effective code review still remains a challenge,
and \emph{identifying appropriate reviewers} for the code review is even more challenging.
To date, both reviewer selection and code review are performed manually at GitHub.
Our tool recommends appropriate developers (\eg\ \emph{cgooding-va}) for such code review task (\eg\ Fig. \ref{fig:mcr}) at GitHub.

\section{CORRECT: Proposed Tool}\label{sec:correct}
Fig. \ref{fig:ui} shows the user interface of CORRECT where we contribute in (d) browser's tool bar, (e)-(f) recommendation panel, and (g) pull request body panel.     
This section discusses different technical features provided by our tool.

\textbf{(1) Use Cases:} CORRECT provides automatic recommendation supports for two use cases of pull request based collaborative software development as follows:

\textbf{(i) Submission of a New Pull Request:} 
During the submission of a \emph{new pull request}, a user (\ie\ developer) compares her project branch (\eg\ \texttt{AA-2453, Fig. \ref{fig:ui}-(a)}) with the base repository (\eg\ \texttt{develop}, Fig. \ref{fig:ui}-(a)), and then looks for potential code reviewers. 
Our tool analyzes the changed source code files (\eg\ Fig. \ref{fig:ui}-(b)) using static analysis, and then suggests a ranked list of appropriate code reviewers in the recommendation panel (\eg\ Fig. \ref{fig:ui}-(e)).     

\textbf{(ii) Update of an Existing Pull Request:} CORRECT can also recommend code reviewers for an \emph{already submitted pull request} for which either no reviewers were assigned or inappropriate reviewers were assigned.  
This could be really helpful since inappropriate assignment of code reviewers often costs precious development time \cite{pick}.   
Our tool analyzes the changed source files from the pull request using static analysis, and then suggests the code reviewers.
% (Fig. \ref{fig:ui}-(e)).     

\textbf{(2) Automatic Mining of Relevant Artifacts:} Our tool automatically mines both version control history and code review history of a software project for identifying appropriate code reviewers for a given pull request as follows:

\textbf{(i) Analysis of Changed Source Code Files:}  Once either a branch (\ie\ first use case) or a pull request (\ie\ second use case) is selected, the tool collects all the changed files from each of their commits from the version control history. 
It accesses GitHub API end points for collecting the changed files, and uses \texttt{github-api}\footnote{https://github.com/kohsuke/github-api}, a popular GitHub client library for the API access.
Since each request or branch might involve a number of source code files, the tool only collects the path of the changed files from the API, and then applies that information to a local mirror of the GitHub repository for performing further static analysis.

\textbf{(ii) Analysis of Code Review History:} Our tool learns to recommend from past code reviews as  was also learnt by existing literature \cite{pick,xin,reduce}. 
It thus collects code review details of the 30 most recently submitted pull requests using GitHub API \cite{correct}. Since online API access could be time-consuming and could hurt the tool's performance, we adopt parallelization in the API access.  
In particular, we apply Java multi-threading to API access and further analysis for each of the past pull requests from the history.
  
\textbf{(3) Automatic Recommendation of Code Reviewers:} CORRECT returns a ranked list of five code reviewers for any given pull request (\ie\ Fig. \ref{fig:ui}-(e)). The size of this recommendation is customizable, and the recommendation generally takes 10-15 seconds on average.
The tool also provides additional insights to assist the user (\ie\ developer) in the selection of appropriate code reviewers for her pull request.  
In particular, it provides relative expertise estimates (\ie\ estimated by our original technique \cite{correct}) of the recommended reviewers on the \emph{external software libraries} and \emph{specialized technologies} used by the pull request.  
Our tool also provides several usability features as follows:

\textbf{(i) Automated Use of Recommendation:} Once code reviewers are recommended in the recommendation panel, the user can copy and paste the reviewers in pull request body panel (\ie\ Fig. \ref{fig:ui}-(g)) by simply clicking \emph{Copy reviewers} button (\ie\ Fig. \ref{fig:ui}-(f)).
Then she can submit the request by using  \emph{Create Pull Request} button (\ie\ Fig. \ref{fig:ui}-(h)).
The tool also provides a \emph{Refresh} button (\ie\ Fig. \ref{fig:ui}-(f)) to help the user start over the reviewer recommendation cycle. 

\textbf{(ii) Caching of Recommendation:} Since we use a stateless protocol--HTTP, caching is a convenient way to improve the performance (\ie\ response time) of the tool. Our tool uses \texttt{localStorage},  a storage feature of Google Chrome and other HTML5-capable browsers,  to store the most recently collected recommendation result.
In the case of repeated requests from the same page (\ie\ branch or pull request), CORRECT displays previously stored result from \texttt{localStorage} database. 
The cache can also be cleared using the \emph{Refresh} button (\ie\ Fig. \ref{fig:ui}-(f)) if the user desires.

\textbf{(4) Personalization \& Optimization:} CORRECT uses open authentication for GitHub API access, and thus, it has the potential not only for personalized reviewer recommendation but also for performance optimization as follows:

\textbf{(i) Personalized Recommendation:} Our tool captures a user's identity from the open authentication step, and then customizes the code reviewer recommendation for her. 
In particular, CORRECT discards self-reference (\ie\ tool user herself as reviewer) from the recommendation list at present. However, other social aspects (\ie\ developer collaboration history) could also be leveraged for further personalization of the reviewer recommendation.

\textbf{(ii) Performance Optimization:} GitHub restricts API access of an average registered user to a rate limit of 5,000 calls per hour. This restriction is likely to introduce \emph{Denial of Service} issue with a tool (\ie\ accessing GitHub API) 
if it is confined to one user account only, especially in an industrial context that involves frequent API access.  
%is deployed in an industrial context
Our tool overcomes that challenge using open authentication where the tool accesses the GitHub API on behalf of the logged in tool user, and thus, the access rarely exceeds the rate limit. 

\vspace{.2cm}
\section{Working Methodology}\label{sec:methodology}
\begin{figure}[!t]
\centering
\includegraphics[width=3.3in]{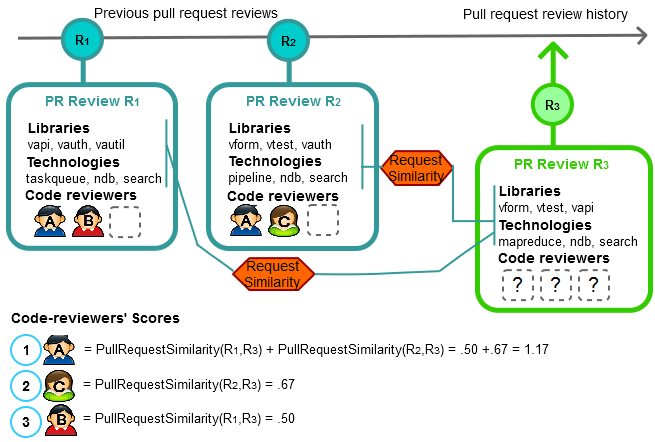}
\vspace{-.6cm}
\caption{Working methodology of CORRECT, taken from the original paper \cite{correct} }
\vspace{-.5cm}
\label{fig:correct}
\end{figure}

Fig. \ref{fig:correct} shows the schematic diagram of our proposed tool-- CORRECT.
Our tool analyzes both version control history and code review history of a software project, and then suggests a ranked list of potential code reviewers for any given pull request.
This section discusses the internal structures and working methodologies of the tool in brief while we refer the readers to the original paper \cite{correct} for further details.  

\textbf{Working Modules:} CORRECT adopts client-server architecture, and it has two working modules--(1) \emph{recommendation engine} and (2) \emph{client module}. 
We package the \emph{client module} as a Google Chrome plug-in and the \emph{recommendation engine} as a web service.          
Once the plug-in is installed successfully, it appears as a \emph{user icon} at the web browser's tool bar (\eg\ Fig. \ref{fig:ui}-(d)).
While the plug-in captures the technical details of a pull request from the web browser, the web service analyzes both the request and other relevant artifacts from the histories, and derives code reviewer recommendation for the request. Both modules communicate using REST and AJAX on top of HTTP.

\textbf{Historical Data Collection:}
CORRECT collects 30 past \emph{CLOSED} pull requests and their corresponding review details from a project for recommending code reviewers for a new pull request.
We first identify each of those pull requests and extract their corresponding commits. Each of these commits can be identified using their SHA-1 based ID, and they generally contain one or more source files that were changed together. 
We collect such changed files from each of the selected past pull requests using GitHub API access and local repository analysis. 
We repeat the same steps for the new pull request, and collect the changed source files to be submitted to the base repository.

We then analyze the code review details of each of the past pull requests, and collect their corresponding reviewers using GitHub API access.
In particular, we collect both the reviewers who were referred to during the submission (\eg\ \emph{rwiebe-va}, Fig. \ref{fig:ui}-(g)) and the reviewers who actually reviewed the pull request (\eg\ \emph{cgooding-va}, Fig. \ref{fig:mcr}).
Such historical information provides the foundation (\ie\ ground truth) for the learning and evaluation of our tool.    

\textbf{Code Review Skill \& Reviewer Ranking:}
Our key idea is-- the developers who have reviewing experience on similar (\ie\ relevant) past pull requests are suitable candidates for reviewing the current pull request to be submitted \cite{pick,reduce}.  
Once changed source code files and review details from the past pull requests are collected, we determine their relevance to the current request based on their shared external libraries (\eg\ \texttt{vapi, vform}) and adopted specialized technologies (\eg\ \texttt{taskqueue, ndb}) in the changed files. 
In particular, we extract the external library or specialized technology names from each pull request, and determine \emph{cosine similarity} between the current request and each of the past requests.  
We then propagate the similarity estimates (as a proxy to review expertise) to the corresponding code reviewers of the past requests. 
Thus, according to CORRECT, the software developers who have more experience on the attached external libraries (\ie\ cross-project experience) and the adopted specialized technologies in the changed files of the current pull request, are more appropriate for the code review than the ones having less experience.

\textbf{Example:} 
Let us consider $R_3$ (Fig. \ref{fig:correct}) is a pull request to be submitted, and the submitter is looking for one or more
code reviewers for the request. $R_1$ and $R_2$ are two past requests similar to $R_3$ containing one or more changed files.
From Fig. \ref{fig:correct}, we see that each of $R_1$ and $R_2$ includes three libraries, adopts three specialized technologies, and is reviewed by a different set of developers. Similarly, $R_3$ also includes three external libraries and adopts three specialized technologies in the changed files.
In order to recommend reviewers for $R_3$, CORRECT first determines the \emph{cosine similarity} between libraries and technologies of $R_3$ and those of $R_1$ and $R_2$.
It then applies those scores to the corresponding reviewers of $R_1$ and $R_2$.
Thus, the developers who have the most review experience with similar past requests, bubble up in the ranked list for code reviewers.
From Fig. \ref{fig:correct}, we see that reviewer $A$ scores the top (\ie\ 1.17) within all the reviewers according to our ranking algorithm, and thus, reviewer \emph{A} is recommended as the code reviewer for the current request--$R_3$.
We recommend the top five code reviewers \cite{reduce,pick} from such a ranked list for any given pull request.
%in order to avoid old but similar requests for reviewer recommendation.   

\section{A Use Case Scenario}
By means of a use case scenario, we attempt to explain how our tool--CORRECT--can help a software developer in choosing appropriate code reviewers for her pull request from within the context of a web browser.

\begin{figure}[!t]
\centering
\includegraphics[width=3.3in]{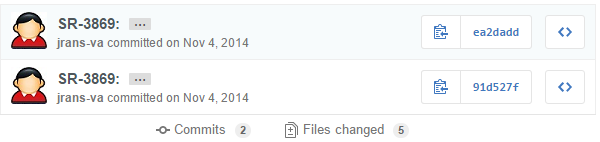}
\vspace{-.6cm}
\centering
\caption{Example use case: Commits \& changed source code files (to be submitted)}
\vspace{-.4cm}
\label{fig:commit}
\end{figure}

Suppose, a developer, \emph{Alice}, has started to collaborate on a new software project --\texttt{SR}-- of Vendasta Technologies. She first forks from the base project which provides her a local copy of the project with complete access for code editing and committing.
She then starts to fix a reported bug with ID-- \texttt{SR-3869}-- where she deletes 28 lines of code and adds 26 lines of code to the local project.
When she is done with the fixation, it is found that the changes were made to five source code files bundled into two commits (\ie\ Fig. \ref{fig:commit}). 
Then she attempts to submit the changes to the base repository using a pull request. 
Modern software companies like Vendasta often have a mandatory requirement for code review in order to maintain the code quality.
Hence, she is also concerned about submitting the changes of higher code quality.
During the pull request submission, she thus attempts to choose a list of expert developers who would review the changes before accepting them as a contribution to the base project. 
To date, GitHub does not provide any support for this task, and thus, she faces several challenges at this stage--(1) Who is the most appropriate code reviewer for these changes? (2) How to determine the code review skill of a developer? and (3) Can we possibly identify appropriate reviewers from the past code reviews or version control history after all? 

She might consider the original authors of the changed files as reviewer candidates. However, this might not be practical since the changed files might be authored by a number of developers over the years who might not be even with the company anymore.
For this use case, we note that nine developers authored the changed files. \emph{Alice} still needs to identify the most appropriate reviewers from those authors by herself with little or no helpful insights about them, which is a challenging task. 
The task is even more challenging for \emph{Alice} if she is novice and/or non-familiar with the fellow developers or the code repositories of the company.

\begin{table}[!t]
\centering
\caption{ Libraries \& Technologies of Use Case}
\label{table:libtech-uc}
%\vspace{-.2cm}
\resizebox{3.3in}{!}{%
\begin{threeparttable}
\begin{tabular}{l|l}
\hline
\textbf{External Library} & \textbf{Specialized Technology} \\
\hline
\hline
\texttt{vapi, vtax, vbcsdk,} & \texttt{google.appengine.ext,}\\
\hhline{~~}
\texttt{vautil, vbcsdk.keys,} & \texttt{ndb, search,}\\
\hhline{~~}
\texttt{vautil.validators.email} & \texttt{google.appengine.api.search}\\
\hline
\end{tabular}
\center
%$^1$Top-K Accuracy
\end{threeparttable}
%\vspace{-.2cm}
}
\vspace{-.6cm}
\end{table}

Now, let us assume that \emph{Alice} has installed CORRECT plug-in on her Chrome web browser, and she initiates a pull request for submission.
Our tool automatically collects the changed source files from the forked project using GitHub API access. Then it suggests her a ranked list of potential code reviewers by analyzing the most recently submitted similar pull requests (\ie\ with code reviews) from the version control history of the base project. 
In particular, the tool automatically extracts \emph{external library} and \emph{specialized technology} information (\eg\ Table \ref{table:libtech-uc}) from each of the changed source files from each pull request, and then leverages the extracted information for code reviewer recommendation (Section \ref{sec:methodology}).
%considers the \emph{external libraries} and the \emph{specialized technologies} used in the code  for determining relevance (\ie\ similarity) between the past pull requests and the current request to be submitted.
Besides ranking, the tool also provides additional insights on the library and technology related experience of the code reviewers (\ie\ Table \ref{table:revs}) which help \emph{Alice} choose the right reviewers.
For example, the top three reviewers--\emph{cberenik-va, cgooding-va} and \emph{ywang-va}--in the recommended list have the maximum expected experience, and she can confidently choose them as the reviewers for her changes. 
%We investigate the code reviews for the branch--\texttt{SR-3869}--from the review history, and found that \emph{cberenik-va} and \emph{cgooding-va} actually reviewed the code for \emph{Alice}.
Thus, to overcome the challenges she faced previously, our tool (1) automatically suggests her a ranked list of appropriate code reviewers for the pull request (to be submitted), (2) automatically captures and leverages external library experience as well as specialized technology experience of the developers as suitable proxies to their code review skills, and (3) automatically mines both version control history and code review history using GitHub API access for deriving the recommendation.
In short, our tool does all the heavy lifting for \emph{Alice} in the background, and she can just get the recommendation by simply clicking a button during the pull request submission. 
More interestingly, CORRECT provides the recommendation within the context of the web browser which helps her maintain the usual work flow (\ie\ within GitHub) and avoid the unexpected context-switching. 
In the context of Vendasta Technologies,  we chose Google Chrome as the web browser. However, any browser plug-in capable of HTTP access can easily consume our recommendation service.       

\begin{table}[!t]
\centering
\caption{Recommended Reviewers for Use Case}
\label{table:revs}
%\vspace{-.2cm}
\resizebox{3.3in}{!}{%
\begin{threeparttable}
\begin{tabular}{l|c|c|c}
\hline
\textbf{Reviewer} & \textbf{Total Score} & \textbf{Library Score} & \textbf{Technology Score}\\
\hline
\hline
\emph{cberenik-va} & 100\% & 64\% & 100\%\\
\hline
\emph{cgooding-va} & 99\% & 100\% & 99\%\\
\hline
\emph{ywang-va} & 75\% & 52\% & 76\%\\
\hline
\emph{sgryschuk-va} & 8\% &  17\% & 8\%\\
\hline
\emph{ksookocheff-va} & 6\% & 0\% & 6\%\\
\hline
\end{tabular}
\center
%$^1$Top-K Accuracy
\end{threeparttable}
%\vspace{-.2cm}
}
\vspace{-.6cm}
\end{table}

\section{Evaluation \& Validation} \label{sec:experiment}
One of the most effective ways for evaluating a code reviewer recommendation system is to consult with actual code reviews and the reviewers assigned for them from a codebase.
%We evaluate our technique using 13,081 pull requests and their code review details from VendAsta codebase. 
%We use a sliding window based selection of requests  where a sliding window approach is used on pull request history for evaluation.
%In order to validate our technique, we also compare with one state-of-the-art technique, and experiment with six open source projects from GitHub.
We evaluate our technique using the real code review  data from Vendasta codebase. 
In particular, we use 13,081 pull requests and their code review details from Vendasta as our oracle in evaluating CORRECT against a number of popular performance metrics.
In order to further validate our findings and demonstrate its superiority, we experiment using 4,034 pull requests from six open source systems of three different programming languages, and compare with the state-of-the-art technique. 
%While the tool has been empirically evaluated and validated, we also plan to validate its applicability through a user study.
While we discuss our evaluation and validation in brief as follows, the details can be found in the original paper \cite{correct}.

%We particularly answer the following research questions through our conducted experiments: 
%\begin{itemize}[noitemsep,topsep=1pt]
%%\item \textbf{RQ$\mathbf{_1}$:} Can relevance between two pull requests in terms of shared external libraries and specialized technologies be exploited in code reviewer recommendation?
%\item \textbf{RQ$\mathbf{_1}$:} How does our technique--CORRECT perform in terms of the state of the art performance metrics?
%%\item \textbf{RQ$\mathbf{_2}$:} Are the \emph{cross-project (\ie\ external library) experience} and the \emph{specialized technology experience} useful proxies for code review skills? 
%\item \textbf{RQ$\mathbf{_2}$:} Does CORRECT outperform the state of the art technique for reviewer recommendation?
%\item \textbf{RQ$\mathbf{_3}$:} Does CORRECT perform equally on both private and public codebase?
%\item \textbf{RQ$\mathbf{_4}$:} Does CORRECT show bias to any of the development frameworks?
%\end{itemize}

\textbf{Evaluation using Vendasta Systems:}
We evaluate our recommendation technique \cite{correct} using a collection of 13,081 pull requests from 10 subject systems (of Vendasta Technologies) and four state of the art performance metrics-- Top-K Accuracy, Mean Reciprocal Rank, Mean Precision and Mean Recall.
CORRECT provides a Top-5 accuracy of 92.15\% and a Mean Reciprocal Rank of 0.67 with 85.93\% precision and 81.39\% recall which are highly promising according to relevant literature \cite{pick,reduce,xin}.
%Mean Reciprocal Rank of 0.67 which is quite high \cite{reduce}. 
%More interestingly, on average, the recommendation technique returns results with 85.93\% precision and 81.39\% recall which suggests its greater potential for recommendation.

\textbf{Comparison with the State-of-the-Art:}\label{sec:compare}
We validate the performance of our technique by comparing with \citet{pick}, the state-of-the-art technique for code reviewer recommendation which outperformed the earlier techniques. Our technique--CORRECT-- provides 11.43\% improvement in Top-5 accuracy and about 10\% improvement in both precision and recall over the state-of-the-art. Three statistical tests-- \emph{MWU, Cohen's d} and \emph{Glass $\bigtriangleup$} -- also suggest that such improvements are statistically significant.

 %Thus, each of the analyses above shows that our technique outperforms the state of the art--RevFinder which was found to be superior to earlier techniques \cite{pick}. This clearly answers \textbf{RQ$\mathbf{_2}$}. 
%Our experimental results suggest that library or technology information is probably more effective than source file path \cite{pick} for pull request relevance which probably led to our better performance.
%
%One might wonder why we did not compare with another relevant technique-- ReviewBot \cite{reduce} from the literature that exploits line change history of source code for reviewer recommendation.
%We made that choice due to two appealing reasons. 
%First, RevFinder outperforms ReviewBot by a large margin. Second, 70\%--90\% of the source code lines in the project are generally changed only once \cite{pick}. 
%Thus, there is an inherent lack of sufficient line-level history, and therefore, the performance of ReviewBot is limited.

%Another recent work--\citet{ghadeer} applies code clone detection (\ie\ code similarity) in expert recommendation for programming problem solving, and is related to our work technically.
%However, our investigation reports that on average only 6.60\% of the source files from each of the subject systems contains code clone segments.
%This suggests that the pull requests also contain very little similar code segments, and thus, source code level similarity between a new request and the past requests might not be effective enough to identify the appropriate code reviewers for the request.

\textbf{Experiments with Open Source Projects:}
Although CORRECT was sufficiently evaluated using \emph{Python} systems from Vendasta, 
we conduct another experiment with six open source projects from GitHub written in three different programming languages--\emph{Java, Python} and \emph{Ruby}-- to generalize our findings.
In this case, CORRECT recommends with a Top-5 accuracy of 85.20\%, a Mean Reciprocal Rank of 0.69, a Mean Precision of 84.76\% and a Mean Recall of 78.73\%.
Comparison demonstrates that our technique outperforms the state-of-the-art \cite{pick} with statistically significant margin.
Further investigations also confirm that CORRECT does not show bias to any programming languages or any project types--\emph{open source} and \emph{closed source}.

\textbf{Evaluation Plan with User Study:} While CORRECT is found promising based on empirical evaluation, we plan to evaluate the tool using a user study involving professional developers from Vendasta. The goal of the study is to determine the usability and usefulness of the tool based on actual developers' feedback. In the user study, we plan to involve at least 10 developers working on 10 different running projects. 
Each participant will install the tool, use it for two controlled tasks (\ie\ code reviewer assignment), and then will evaluate the recommendation provided by the tool with a predefined rating scale. We would then collect the numerical ratings as well as their qualitative feedback to triangulate them with our empirical findings.

\section{Related Work} \label{sec:related}
\textbf{Code Reviewer Recommendation}: Existing studies recommend code reviewers by analyzing mostly code review or version control history \cite{yu, xin,rveffect,pick,reduce} and developer collaboration networks \cite{yu}.
%-- line change history \cite{reduce} and past review comments \cite{yu, xin}, project directory structure \cite{pick, rveffect}, and developer collaboration network \cite{yu}.  
%employ different techniques such as version control history \cite{reduce, kevic, bosu}, directory structure \cite{rveffect} and authors' meta data \cite{authorship, discovery} analysis, code dependency \cite{brazil} and code similarity \cite{ghadeer} analysis  and developer communication history \cite{brazil, seafood, sna, ghadeer} analysis. 
%In summary, these approaches focus on the analysis of source file level artifacts or interactions among the developers for determining the expertise of a developer.
\citet{reduce} proposes \emph{ReviewBot} that analyzes \emph{line change history} of the affected source lines from a given review request, and then identifies code reviewers from that history for the request.
However, existing findings suggest that most of the lines are generally changed only once \cite{pick} which makes the line change history really scarce and thus, the performance of  ReviewBot is limited.
\citeauthor{pick} propose \emph{RevFinder} \cite{pick} that identifies relevant review requests using \emph{File Path Similarity (FPS)} \cite{rveffect}, and then recommends reviewers from those requests for a review request at hand.
RevFinder also outperformed earlier techniques including ReviewBot \cite{pick}.
On the other hand, CORRECT identifies relevant pull requests using \emph{external library similarity} and \emph{specialized technology similarity} which are found to be more effective than File Path Similarity\cite{pick} for estimating relevance between pull requests, and thus for reviewer recommendation. 
In our earlier work \cite{correct}, we show that our technique outperformed RevFinder with statistically significant performance improvements.
%We compare with \citet{pick}, a state-of-the-art technique, and experimental results (Section \ref{sec:compare}) show that our technique outperforms it with statistically significant performance improvement.
Another recent work \cite{xin} applies machine learning on past code reviews, and combines textual similarity with File Path Similarity \cite{pick}. Thus, it suffers from similar issues as of RevFinder such as pull request relevance issue, and that the learned models could be biased to the subject systems under study.

%The recent work of \citet{xin} improved upon \citet{pick} by applying machine learning on past review comments and File Path Similarity. Although, we did not directly compare with this technique, it suffers from the same issues of \citet{pick}, and the learned models could be biased to the subject systems under study.

The remaining technique--\citet{yu} analyzes past review comments and developer collaboration networks for reviewer recommendation.
While we use library and technology similarity between pull requests for determining relevant past requests, they use review comment similarity (\ie\ textual similarity) for the same purpose.
Besides, their idea is still not properly evaluated or validated.
%\citet{rveffect} improve this algorithm where they assign reviewer points using \emph{File Path Similarity (FPS)} algorithm.
%According to \citet{rveffect}, developers who reviewed the source files from a project directory of interest are suitable for later code reviews involving that directory.
%Thus both approaches target a subset of past reviews for a later recommendation whereas our approach analyzes three types of contributions-- \emph{authorship, reviewership} and \emph{technology adoption} not only for past review selection but also for code reviewer recommendation. 

\textbf{Expert Recommendation:} \citet{ghadeer} identify expert developers on a code fragment of interest by exploiting \emph{code similarity} with other segments. 
%The baseline idea is, if a developer has worked on code fragments that are similar to the fragment of interest, that developer is a suitable candidate for technically assisting (\eg\ code review) with the fragment. They also apply a list of social heuristics which determine the developer's acceptance in the community, in the recommendation of expert developers. 
Similar technique is applied by \citet{brazil} where they develop a communication network among documents, source code and developers, and then recommend dominant developers as experts. 
%by analyzing \emph{document-source code, developer-source code, source code-source code} and \emph{developer-developer} relationships. %They then locate the dominant developer nodes in the network as the expert developers for recommendation. 
\citet{sna} studies the developer network using \emph{code review relationship}, and identifies core and peripheral developers using different network properties.
%of three open source projects based on code review relationship, and analyze the contribution patterns of core developers and peripheral developers of a project. Thus while these approaches leverage various communication networks for expert developer identification, we adopt a classifier model-based approach for the same task. Furthermore, we consider different aspects of programming contributions in determining expertise for code review.    
There exist several studies in the domain of \emph{bug triaging} that analyze duplicate bug reports \cite{kevic} or apply IR-based traceability \cite{authorship} techniques for recommending appropriate developers for bug fixation.
%developers for bug fixation. \citet{kevic} identify duplicate bug reports of a reported bug by employing text similarity, and analyze the changeset of their fixations for appropriate developer recommendation. Similar approach is adopted by \citet{authorship} who use an IR-based traceability technique to locate relevant source files for a reported bug, and then exploit the authorship meta data for expert developer recommendation.
Several studies are also conducted on expert user recommendation at Stack Overflow that analyze cross-domain contributions \cite{discovery} or question difficulty \cite{difficulty} for expertise estimation.
While these expert recommendation techniques are somewhat similar to ours, their context of recommendation is different and thus, comparing ours with them is not feasible. 
Of course, we introduced two novel and effective expertise paradigms (\emph{cross-project experience} and \emph{specialized technology experience}) which were not exploited by any of the recommendation systems. This makes our proposed tool--CORRECT-- significantly different from all of them.

 %Thus, the above techniques recommend experts in different contexts than ours, and a direct comparison is not required.
%for programming problem solving, bug fixation or question answering,
%In contrast, our technique performs quite a similar job in the context of code review by applying two novel and effective expertise paradigms.
%recommends experts for code review at GitHub.

%Thus, while the above studies focus on recommending experts for programming problem solving, our technique focus on a more specific task such as code review.
%\balance
%\balance

\section{Conclusion} \label{sec:conclusion}
To summarize, we propose a novel tool-- CORRECT-- for code reviewer suggestion for pull requests at GitHub for Vendasta Technologies. 
It automatically captures the experience of a developer with the external libraries (\ie\ cross-project experience) and specialized technologies used in a given pull request, applies such experiences as proxies to code review skill of the developer, and then suggests a ranked list of appropriate code reviewers. 
Our recommendation technique is substantially evaluated and validated using empirical data.
We package our solution as a web service and a plug-in for Google Chrome browser. 
The tool can assist a developer in choosing appropriate code reviewers during the submission of a new pull request or during the update (\ie\ reviewer assignment) of an existing pull request.
\bibliographystyle{plainnat}
\setlength{\bibsep}{0pt plus 0.2ex}
\scriptsize
\bibliography{sigproc}
\end{document}